\documentclass{caosp305}

\usepackage{graphicx}

\usepackage{natbib}
\articleNo{123}
\pubyear{2017}
\volume{?}
\volnumber{3}
\firstpage{1}
\received{November 10, 2017}
\accepted{November 10, 2017}

\newcommand{\aap}{Astron. Astrophys.}
\newcommand{\eprint}[2][]{{\tt #1:#2}}

\begin{document}

\htitle{Surface mapping of magnetic hot stars}
\hauthor{O.\,Kochukhov}

\title{Surface mapping of magnetic hot stars}
\subtitle{Theories versus observations}

\author{
        O.\,Kochukhov
       }

\institute{
           Department of Physics and Astronomy, Uppsala University, SE 75120, Uppsala, Sweden, \email{oleg.kochukhov@physics.uu.se}
          }

\date{November 10, 2017}

\maketitle

\begin{abstract}
This review summarises results of recent magnetic and chemical abundance surface mapping studies of early-type stars. We discuss main trends uncovered by observational investigations and consider reliability of spectropolarimetric inversion techniques used to infer these results. A critical assessment of theoretical attempts to interpret empirical magnetic and chemical maps in the framework of, respectively, the fossil field and atomic diffusion theories is also presented. This confrontation of theory and observations demonstrates that 3D MHD models of fossil field relaxation are  successful in matching the observed range of surface magnetic field geometries. At the same time, even the most recent time-dependent atomic diffusion calculations fail to reproduce diverse horizontal abundance distributions found in real magnetic hot stars. 
\keywords{stars: atmospheres -- stars: chemically peculiar -- stars: magnetic fields -- starspots}
\end{abstract}

\section{Introduction}
\label{intro}

The presence of strong, stable magnetic fields in the outer envelopes of early-type stars leads to formation of prominent horizontal and vertical chemical abundance inhomogeneities. These chemical structures can be mapped by applying Doppler imaging (DI) inversion procedures to high-resolution spectroscopic observations. Moreover, modern time-resolved spectropolarimetric data, in particular high-resolution spectra in all four Stokes parameters, provide sufficient information for reconstruction of detailed geometries of surface magnetic fields, relaxing the common assumption of oblique dipolar magnetic topologies. 

This paper summarises results of recent observational magnetic and chemical abundance mapping studies of hot stars. We also touch upon the question of systematic errors of the surface mapping procedures and their intrinsic limitations. This review is concluded with a critical assessment of theoretical attempts to interpret empirical magnetic and chemical maps of stellar surfaces with the help of ab initio fossil field and atomic diffusion calculations.

\section{Magnetic field mapping}
\label{mdi}

\subsection{Observational results}

Most historic and many recent studies of the magnetic field topologies of hot stars were limited to fitting simple parametrised dipole or dipole plus quadrupole model topologies to the phase curves of integral magnetic observables. With the improvement of observational capabilities and, in particular, introduction of wide wavelength coverage, high-resolution full Stokes vector spectrometers (ESPaDOnS, Narval, HARPSpol) it became possible to reconstruct arbitrary surface magnetic field topologies directly from time-series polarisation profiles with the magnetic Doppler imaging (MDI) technique \citep{piskunov:2002a,kochukhov:2014}. These magnetic inversion studies, carried out for 16 A and B stars so far, demonstrate a key role of linear polarisation data (Stokes $Q$ and $U$ profiles) in detecting and characterising the small-scale field topologies \citep{kochukhov:2004d,kochukhov:2010}. Four Stokes parameter MDI analyses suggest that the typical surface magnetic geometry of an early-type star is represented by a superposition of (occasionally distorted) dipolar fields and local magnetic spots. Without the constraints provided by the Stokes $QU$ data, spectropolarimetric inversions yield a smoother field distribution, almost entirely lacking small-scale details, and may suffer considerable ambiguity in rare cases of strongly non-dipolar field configurations \citep{kochukhov:2016a}. Unfortunately, only 8 ApBp stars have so far been analysed taking the linear spectropolarimetric data into account. 

\begin{figure}[!th]
\centerline{\includegraphics[width=0.70\textwidth,clip=]{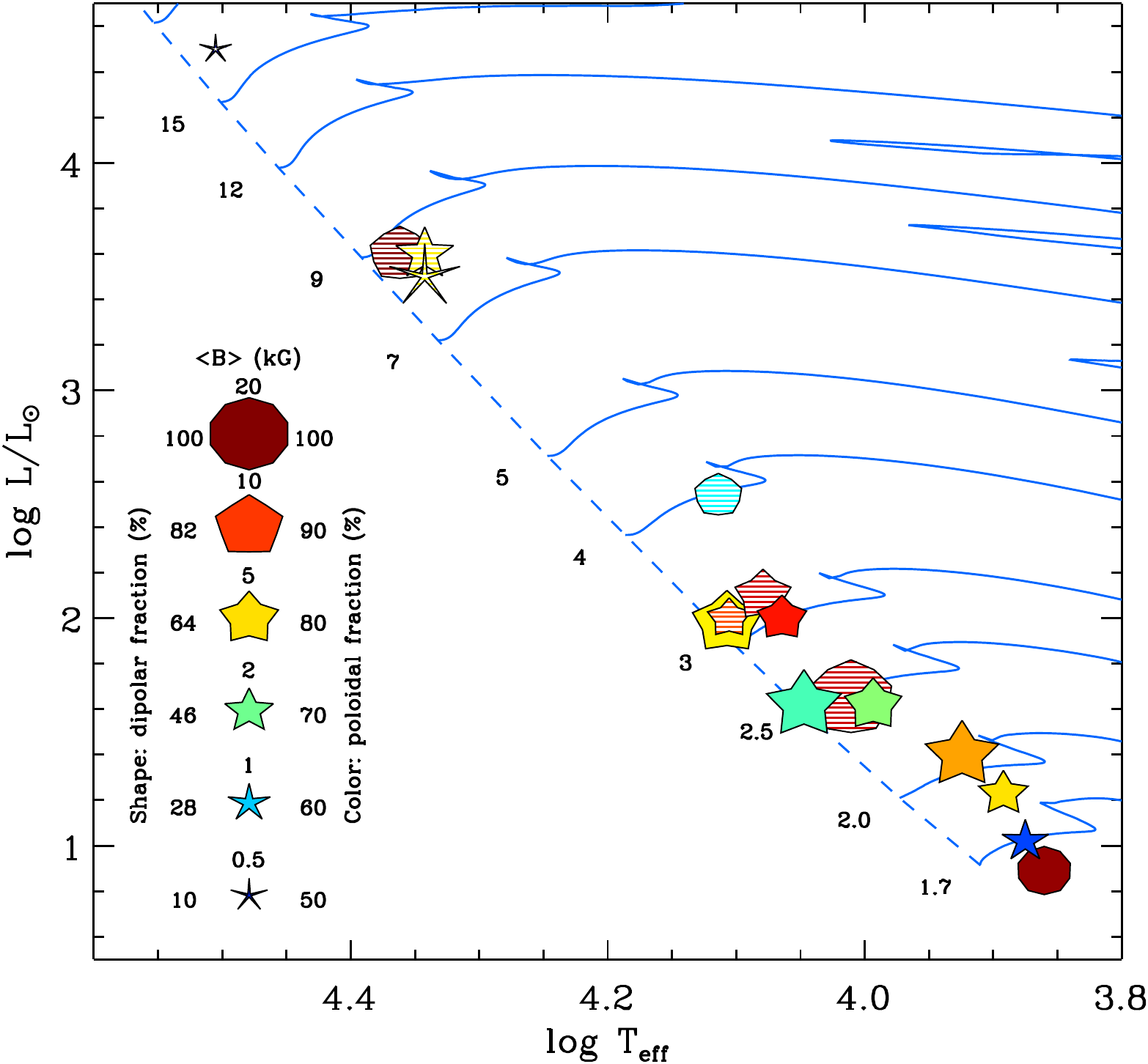}}
\caption{Positions of the hot-star MDI targets on the H-R diagram. The sizes of symbols correspond to the mean magnetic field strength, their shape to the dipolar magnetic energy fraction, and the colour to the ratio of poloidal and toroidal field components. The solid symbols indicate results based on the full Stokes vector inversions while the hatched symbols correspond to the MDI results based on circular polarisation alone.}
\label{mdi_obs}
\end{figure}

Figure~\ref{mdi_obs} presents a summary of all MDI results obtained for early-type magnetic stars. The symbol sizes, shapes and colours encode the average field strength, contribution of the dipolar component and the fraction of toroidal field energy. Generally, the field topologies appear more diverse than thought previously. There are examples of very nearly dipolar fields, even for stars studied with high-quality full Stokes vector data \citep{rusomarov:2015}, clearly non-dipolar and non-axisymmetric field configurations \citep{kochukhov:2011a,kochukhov:2016a}, and field topologies with an unexpectedly large toroidal contribution \citep{oksala:2017}. However, no coherent pattern of the field geometry versus stellar mass or/and age emerges from Fig.~\ref{mdi_obs}, except for the tendency of finding the most complex non-dipolar fields in the more massive stars ($\tau$~Sco, HD\,37776). At the same time, it must be acknowledged that most objects in this figure are field ApBp stars for which the evolutionary status is constrained to no better than about half of the main sequence lifetime.

\subsection{Comparison with theories}

Few theoretical studies went beyond the dipolar approximation in studying the structure of surface magnetic fields in early-type stars. Among more general analyses, MHD simulations of the fossil field relaxation by \citet{braithwaite:2006} and \citet{braithwaite:2008} demonstrated formation of stable, mixed poloidal-toroidal 3D magnetic configurations in stably stratified stellar interiors from initially random seed fields. We have analysed a number of models by Braithwaite (private communication) aiming to compare theoretical surface magnetic field configurations with observational MDI studies of early-type stars. This comparison shows that these theoretical calculations are remarkably successful in encompassing the entire range of the observed field geometries. Similar to empirical magnetic maps, Braithwaite's simulations sometimes produce almost perfect surface dipoles. But in other cases the surface field geometry is a dipole with significant distortions, such as small-scale local spots, or dipolar geometries with a significant toroidal contribution or even strongly non-dipolar fields. However, since these different outcomes of MHD modelling result from adopting different initial conditions in a highly idealised, non-rotating and non-evolving stellar interior model, it is not possible to relate theoretical predictions of different magnetic field geometries with physical properties of real stars or with some particular evolutionary stages.

\section{Chemical abundance mapping}
\label{di}

\subsection{Observational results}

A summary of 39 individual DI and MDI analyses of 36 magnetic Ap/Bp stars and 4 HgMn stars lacking global magnetic fields was complied by \citet{kochukhov:2017}. Five other abundance DI studies were published in 2017. Multi-element (5 or more species) abundance mapping results are available for 15 stars, with HD\,3980 \citep{nesvacil:2012}, HD\,24712 \citep{luftinger:2010}, HD\,50773 \citep{luftinger:2010a}, HD\,83368 \citep{kochukhov:2004e}, $\varepsilon$~UMa \citep{lueftinger:2003}, and $\alpha^2$~CVn \citep{silvester:2014a} studied most comprehensively.

\begin{figure}[!th]
\centerline{\includegraphics[width=0.9\textwidth,clip=]{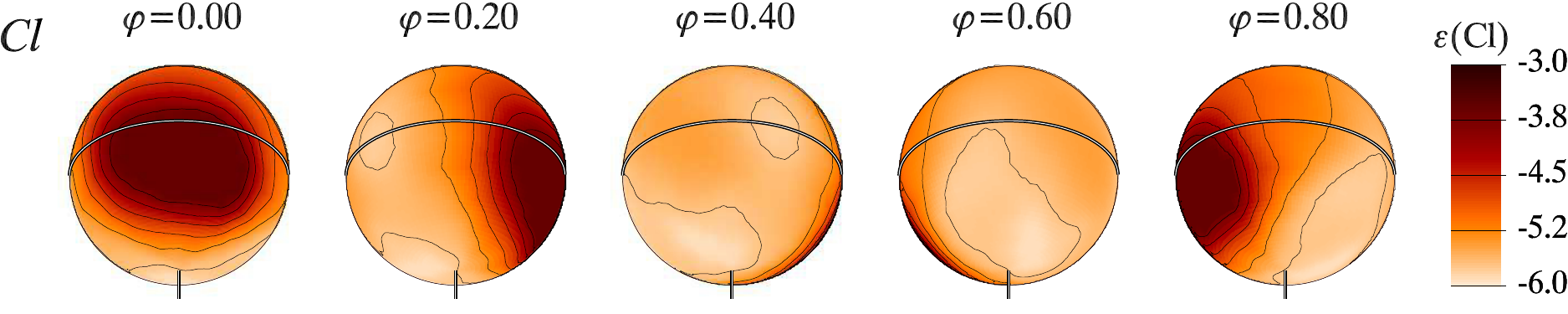}}
\centerline{\includegraphics[width=0.9\textwidth,clip=]{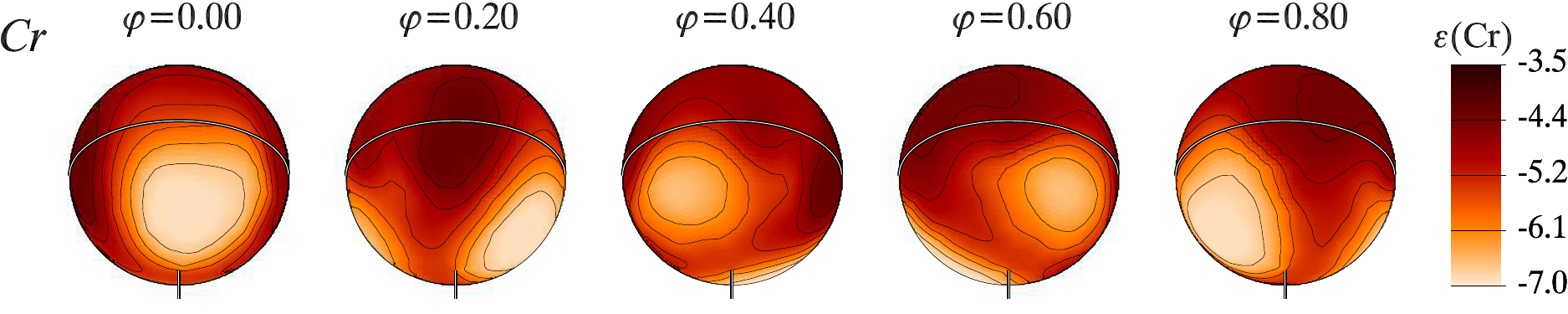}}
\centerline{\includegraphics[width=0.9\textwidth,clip=]{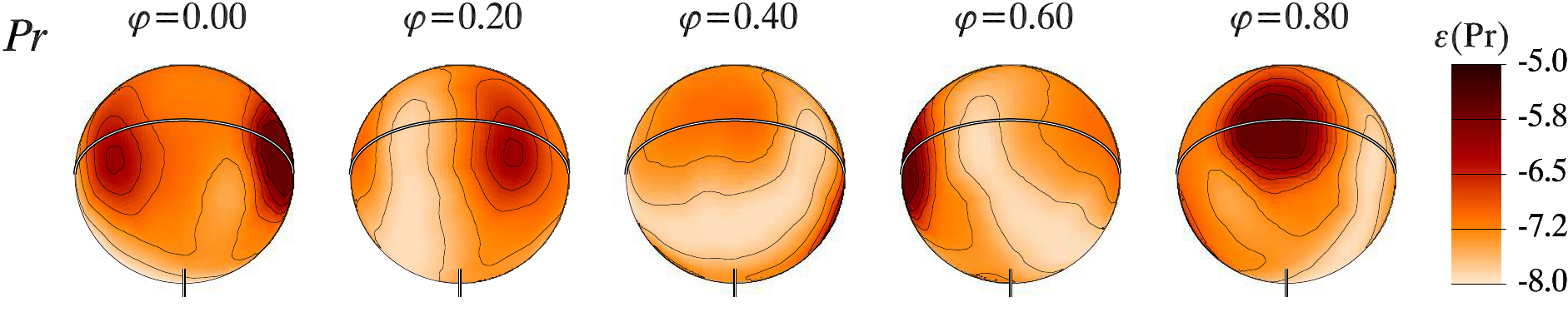}}
\centerline{\includegraphics[height=0.85\textwidth,angle=-90,clip=]{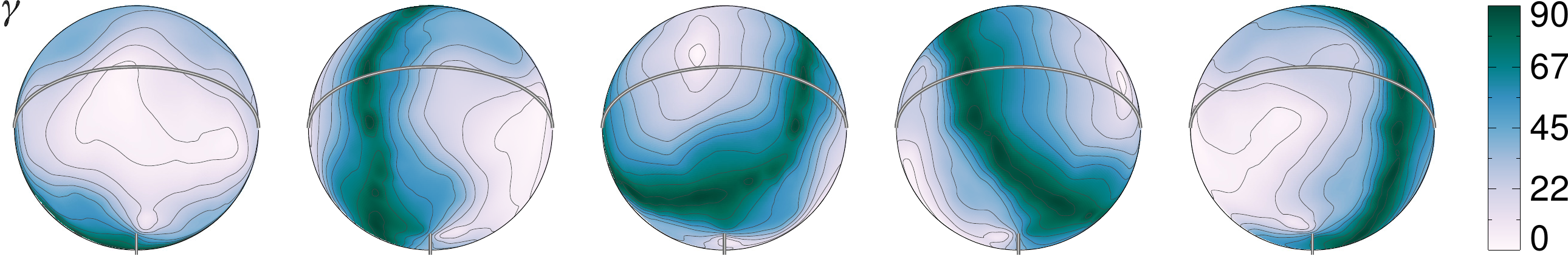}}
\caption{Surface distributions of Cl, Cr, and Pr (top three rows) derived for the magnetic Ap star $\alpha^2$~CVn by \citet{silvester:2014a} in comparison to the local field inclination (bottom row). The element abundances are given in the $\log N_{\rm el}/N_{\rm tot}$ scale. The field inclination is measured in degrees.}
\label{di_obs}
\end{figure}

The published DI maps of early-type stars reveal a striking diversity of surface chemical spot morphologies, both in terms of distributions of different elements for the same star and in terms of behaviour of the same element for different stars with similar atmospheric parameters. It is very common to find elements showing an opposite relation to the magnetic field geometry for the same star. For instance, oxygen appears to be the only example of an element with a systematic enhancement at the equator of dipolar magnetic field \citep{rice:1997,kochukhov:2004e}, while Li, Cl and rare-earth elements tend to concentrate in the vicinity of one or both magnetic poles (see Fig.~\ref{di_obs}). But it is also not uncommon to find significant offsets of major element concentrations from the location of magnetic features (e.g. Pr in Fig.~\ref{di_obs}) or lack of any apparent relation to the field geometry. 

Many recent abundance mapping studies have incorporated detailed treatment of magnetic field effects in polarised radiative transfer and often derived both magnetic field geometry and several chemical distributions with a self-consistent MDI inversion based on observations in several Stokes parameters. However, the bulk of historic Ap-star abundance mapping investigations neglected magnetic fields. Numerical experiments \citep{kochukhov:2017} demonstrate that this leads to average errors of 0.2--0.3~dex in the inferred local abundances for typical dipolar field strength of a few kG. Similar systematic errors arise due to neglect of lateral variations of continuum brightness and model atmosphere structure. These errors are small compared to typical abundance contrasts of 2--5~dex derived with DI for Ap/Bp stars.

Photometric time-series studies provide an independent validation of the spectroscopic DI inversion results. \citet{luftinger:2010a} found that the surface locations of main metal overabundance features in the Ap star HD\,50773 coincide with positions of bright spots derived from the CoRoT light curve of this star. Several other studies \citep[e.g.][]{krticka:2015} compared light curves of Ap/Bp stars in multiple photometric bands with the synthetic SED models based directly on the local abundances inferred by independent DI analyses. These comparisons showed that the amplitudes, shapes and wavelength dependence of the observed photometric phase curves are satisfactorily or, in many cases, very successfully reproduced by multi-element DI maps, confirming validity of abundance inversions over a wide stellar parameter range.

\subsection{Comparison with theories}

The origin of chemical spots in early-type stars is currently not understood. It is generally believed that anomalous surface chemistry of Ap/Bp stars is explained by the selective gravitational settling and radiative levitation of atoms, the process known as atomic diffusion \citep{michaud:2015}. Non-magnetic, spherically symmetric diffusion calculations in stellar interiors \citep{vick:2010} and atmospheres \citep{leblanc:2009} are indeed reasonably successful in reproducing the observed temperature-dependent mean element abundances and mean vertical stratification of chemical elements \citep{ryabchikova:2011}. 

In comparison, little progress is evident in the \textit{ab initio} modelling of the horizontal chemical spots. Early, semi-qualitative studies attempting to link spot geometries with an underlying magnetic field structure \citep{michaud:1981,alecian:1981,babel:1991} have been superseded by detailed numerical models taking into account anisotropic diffusion in the presence of an arbitrary magnetic field and including a feedback of chemical stratification on the stellar atmospheric structure \citep{alecian:2010,alecian:2015}. This modelling has been also extended to a time-dependent treatment of diffusion \citep{alecian:2011,stift:2016}, lifting the common arbitrary assumption of an equilibrium element stratification (zero net particle flux). At the same time, these diffusion calculations neglect several effects (ambipolar diffusion, weak stellar wind, electromagnetic particle drifts, realistic bottom boundary condition provided by the interior particle diffusion) already shown to be important by previous studies. None of the recent diffusion studies have considered stellar rotation and related global circulation and mixing effects.

Probably as a result of one or several of these simplifications, theoretical horizontal abundance distributions fail to achieve even a qualitative agreement with observations. Most importantly, both earlier equilibrium and more recent time-dependent calculations predict all chemical elements to be either homogeneously distributed or share the exact same distribution characterised by a preferential element concentration in (very nearly) horizontal field regions. As illustrated by Fig.~\ref{di_obs} for $\alpha^2$~CVn as well as by other recent DI results and many historic spectrum variability studies, this conjecture is clearly at odds with reality (with a possible exception of oxygen distributions in a couple of Ap stars).

\begin{figure}[!t]
\centerline{\includegraphics[height=0.9\textwidth,angle=-90,clip=]{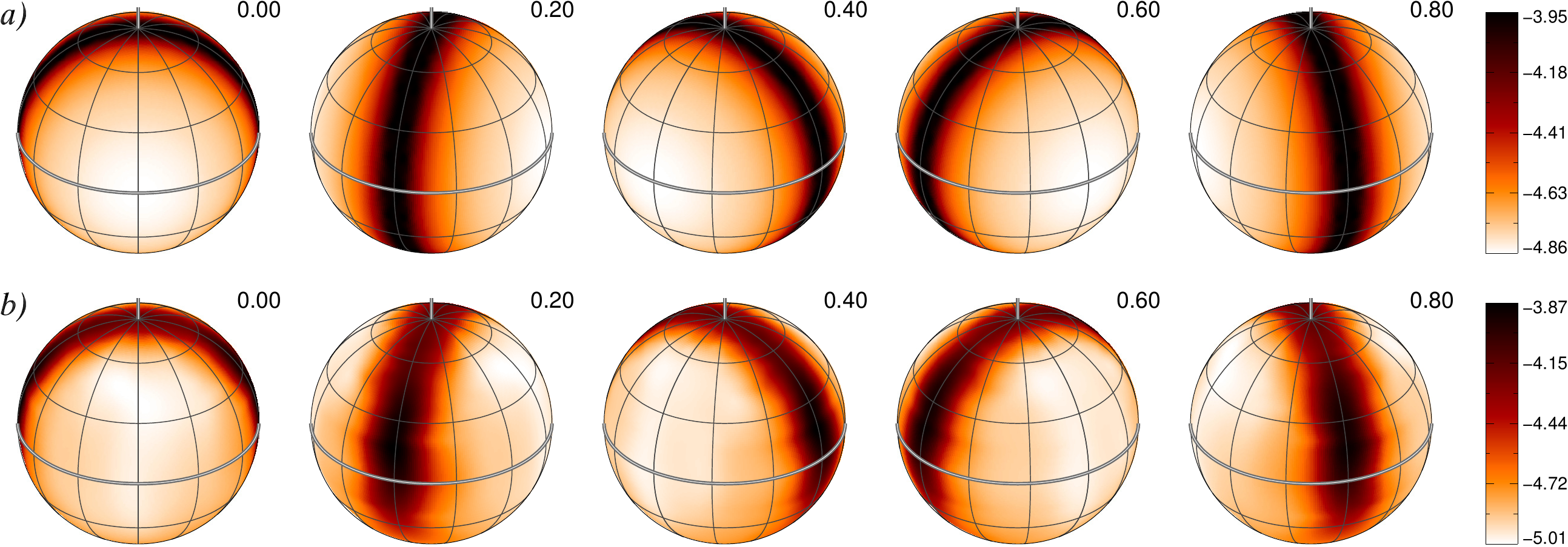}}
\caption{Horizontal distribution of Fe abundance at the optical depth $\log \tau_{5000}=-2$ from the theoretical 3D diffusion calculations by \citet{stift:2016} (a) compared to the 2D Doppler imaging reconstruction from simulated data for four Fe~{\sc ii} lines (b).
}
\label{di_test}
\end{figure}

A detailed analysis (Kochukhov \& Ryabchikova, submitted) of the 3D Fe and Cr theoretical chemical profiles published by \citet{stift:2016} reveals that these time-dependent diffusion calculations underestimate the mean abundances of these elements by a factor of 100--1000, thus, paradoxically, leading to a far worse agreement with the observed mean abundances compared to, presumably less realistic, equilibrium diffusion models of previous generation.

Observational spectroscopic data simulated according to the diffusion model by \citet{stift:2016} can be used to test the impact of vertical chemical stratification (neglected by typical horizontal mapping studies) on the DI results, thereby assessing another, potentially important, systematic error source. The outcome of this experiment, illustrated in Fig.~\ref{di_test}, proves that neglect of vertical chemical stratification does not lead to a noticeable deterioration or a systematic bias in the DI maps. The usual 2D inversions are able to recover horizontal element distributions that closely match a representative cross-section of the input 3D abundance map. These results confirm that lateral cross-sections of the 3D diffusion model predictions can be meaningfully compared to the entire body of published 2D DI maps.

\section{Conclusions and outlook}
\label{concl}

Modern spectral inversion techniques are able to provide detailed and reliable maps of chemical element and magnetic field vector distributions on the surfaces of Ap/Bp and related stars. Much of the recent progress in understanding magnetic field geometries of these objects is spurred by the extension of spectropolarimetric magnetic studies from the traditional circular polarisation diagnostic to a full Stokes vector analysis. However, the number of studies based on four Stokes parameter spectra needs to be increased in order to obtain an unbiased view of early-type star magnetism.

The evolutionary perspective has been almost entirely missing from the surface mapping studies owing to their focus on bright, easily accessible field stars. Systematic DI and MDI analyses of cluster Ap/Bp stars are necessary for putting surface structure information in a relevant evolutionary context.

The current theoretical atmospheric atomic diffusion calculations are fundamentally unable to account for the observed diversity of chemical spot distributions in Ap/Bp stars. The latest time-dependent diffusion calculations also fail to reproduce a large mean overabundance of Fe-peak elements, known to be a universal distinguishing feature of most of these objects. It is likely that one or several physical processes (e.g. weak mass loss, stellar rotation, electromagnetic particle drifts) ignored by the current diffusion calculations are playing important roles in shaping non-uniform element distributions in early-type stars.

\acknowledgements
This research is supported by the Knut and Alice Wallenberg Foundation, the Swedish Research Council, and the Swedish National Space Board. The author thanks Dr. J. Braithwaite for providing MHD models of fossil magnetic fields and Drs. G.A.~Wade, J.~Silvester and N.~Rusomarov for many stimulating discussions and contributions to the studies described in this review.


\end{document}